\documentclass[twocolumn,english,aps,prl,floatfix,]{revtex4}

\usepackage[T1]{fontenc}
\usepackage[latin1]{inputenc}

\makeatletter

\usepackage{babel}

\usepackage{babel}
\makeatother

\begin{document}

\title{Limitations of spin-fermion models in studying underdoped cuprates}

\author{Alvaro Ferraz$^{1}$, Evgenii Kochetov$^{1,2}$}

\affiliation{$^{1}$International Institute of Physics - UFRN,
Department of Experimental and Theoretical Physics - UFRN, Natal, Brazil}
\affiliation{$^{2}$Bogoliubov Theoretical Laboratory, Joint
Institute for Nuclear Research, 141980 Dubna, Russia}



\begin{abstract}
A microscopic basis is provided for the spin-fermion model used to describe the physics of the underdoped cuprates. In this way, a spin-fermion coupling is shown to take care of the local no double occupancy constraint, which is ignored in the weakly coupled regime. This ingredient is proved to be essential to capture the physics of strong correlations, however. We elaborate further on our recent proposal for the strong-coupling version of the spin-fermion model that may prove to be the correct starting point to deal with the strong correlation physics displayed by the underdoped cuprates.
\end{abstract}

\maketitle

\section{I. Introduction: Spin-fermion model}

The  purpose  of this work is to establish a reliable microscopic basis for the spin-fermion (SF) model 
which might be appropriate to describe the physics of the lightly doped cuprates. Since there are several different versions of the SF model in use to describe the high-$T_c$ superconductors it is important to discuss initially how they differ from each other.

More than a decade ago the SF model was put forward in an attempt to account for 
the observed non-Fermi liquid anomalies in the doped cuprates near optimal doping.\cite{chub} 
This original model essentially describes low-energy fermions with a large Fermi surface (FS) interacting with each other via soft collective spin excitations,
$${\cal L}_{sf}={\cal L}_{c}+{\cal L}_{\lambda}+{\cal L}_{\phi},$$ where 
\begin{eqnarray}
{\cal L}_{c}&=&\sum_i c^{\dagger}_{i\alpha}(\partial_{\tau}-\mu)c_{i\alpha}-\sum_{i<j}t_{ij}(c^{\dagger}_{i\alpha}c_{j\alpha}+
c^{\dagger}_{j\alpha}c_{i\alpha}),\nonumber\\
{\cal L}_{\lambda}&=&\lambda\sum_i\phi_i^{a} c^{\dagger}_{i\alpha}\sigma^a_{\alpha\beta}c_{i\beta},\nonumber\\
{\cal L}_{\phi}&=&\sum_q \chi^{-1}_0\vec\phi_{\vec q}\vec\phi_{-\vec q}.
\label{1.1} \end{eqnarray}
Here  $c^{\dagger}_{i\alpha}$ is the creation operator for electrons with spin projection $\alpha=\uparrow,\downarrow$ on site $i$,  the amplitudes $t_{ij}$ are hopping matrix elements describing the "large" Fermi surface, $\mu$ is a chemical potential, and $\sigma^a, \, a=x,y,z,$ are the Pauli matrices. 

The bare spin propagator emerges from the integration of the high-energy fermionic modes,
\begin{equation}
\chi_0(q=\vec q,\omega_n)=(\frac{\omega_n^2}{v_s^2}+(\vec q-\vec Q)^2+\xi_{AF}^{-2})^{-1},
\label{1.2}\end{equation}
where $\omega_n=2\pi n/\beta$ are Matsubara bosonic frequencies. The antiferromagnetic (AF) spin correlation length, $\xi_{AF}$, and the spin velocity $v_s$ are high-energy phenomenological parameters. 
At criticality ($\xi_{AF}=\infty$), the singular nature of the low-energy AF paramagnons singles out the so-called {\it hotspots}
on the FS connected by $\vec Q$. Those points are precisely the intersects of the large 
hole-like FS with the magnetic zone boundary.

Due to the enormous  enhancement of the AF fluctuations at or near  the quantum critical point (QCP), the fermions in the vicinity of the hotspots interact strongly with each other via singular AF paramagnons. This results in strong damping of the propagating spin modes.  The spin modes become dissipative and the fermions quasiparticles  acquire a short lifetime. As a result, the electron system displays a singular non-FL infrared behaviour declared to be observed in the cuprates near optimal doping.\cite{chub} 
In principle, the SF model is a FL theory plagued with infrared singularities 
in an attempt to account for a non-FL low-energy physics of the optimally doped cuprates.

This SF model can be formally derived from the Hubbard model \cite{chub} which describes fermions hopping on a lattice with a tunneling amplitude $t_{ij}$ and are subjected to a short-range Coulomb repulsion $U$,
\begin{equation}
H_U=-\sum_{i,j}t_{ij\alpha}(c^{\dagger}_{i\alpha}c_{j\alpha}-\mu)+U\sum_in_{i\uparrow}n_{i\downarrow},
\label{1.3}\end{equation}
where $n_{i\alpha}=c^{\dagger}_{i\alpha}c_{i\alpha}$ is the on-site electron number operator.
Following standard procedure the quartic fermion interaction can be decoupled using a Hubbard-Stratonovich field $\vec\phi$ that physically corresponds
to the collective spin excitations which are "made out" of the fermion degrees of freedom. This procedure can be justified only if the 
SF coupling $\lambda\sim U$ is smaller that the fermionic bandwidth $W\sim t$, where $t$ is a nearest-neighbour hopping amplitude.
This automatically brings the theory into  a weak-coupling regime controlled by a small parameter, $U/t$. 

The requirement  $U/t\ll 1$ implies that strong electron correlations due a large on-site Coulomb repulsion $U$ are totally ignored. This may correspond to an overdoped region of the cuprate phase diagram that displays a conventional FL physics. With some modifications (the hot spots theory, etc.) being imposed, it exhibits a non-FL behaviour that might be thought of  as being relevant to the optimally doped cuprates, provided the strange metal phase has nothing to do with strong correlations. However, this approach is inapplicable to lower dopings in which strong correlations are known to be at work. In particular, the underdoped pseudogap (PG) region is characterized by strong electron correlations which keep the majority of the electrons well localized. Accordingly, the FS consists of small separated patches (electron pockets) as observed in the quantum-oscillation experiments.

The observation of quantum oscillations in the lightly hole-doped cuprates is an important breakthrough since it indicates that coherent electronic quasiparticles 
may exist even in the PG regime. The PG state does not exhibit a large FS enclosing a total number of charged carriers. Instead
the FS consists of small pockets with a total area proportional to a dopant density $x$, rather than $1+x$ which is expected for conventional FL's.  
There are two routes to a Fermi surface reconstruction that could apply to the cuprates.
A possible theoretical justification for this phenomenon might be the  occurrence  of a simultaneous setting of a new long-range order together with the PG phase 
at a critical hole concentration, $x_c\approx 0.19$. 
The resulting breakdown of translational symmetry would cut the large FS into small pieces but the Luttinger's theorem (LT) would still hold. This is exactly what the conventional SF model predicts.
However, so far there are no  observations of the exhistance of any order parameter that breaks translation symmetry near $x=x_c$. 

\section{II. Fractionalized Fermi-liquid model}

The fractionalized Fermi-liquid (FL$^*$) was proposed phenomenologically as an alternative 
to describe lightly doped antiferromagnets. It goes beyond the weak coupling regime and as a result it is able to produce small Fermi pockets without translational symmetry breaking.
The Lagrangian density of this new model is \cite{sachdev}
$${\cal L'}_{sf}={\cal L'}_{c}+{\cal L'}_{\lambda}+{\cal L'}_{n},$$
\begin{eqnarray}
{\cal L'}_{c}&=&\sum_i c^{\dagger}_{i\alpha}(\partial_{\tau}-\mu)c_{i\alpha}-\sum_{i<j}t_{ij}(c^{\dagger}_{i\alpha}c_{j\alpha}+
c^{\dagger}_{j\alpha}c_{i\alpha}),\nonumber\\
{\cal L'}_{\lambda}&=&\lambda\sum_i(-1)^{x_i+y_i}n_i^{a} c^{\dagger}_{i\alpha}\sigma^a_{\alpha\beta}c_{i\beta},\nonumber\\
{\cal L'}_{n}&=&\frac{1}{2g}\int d^2r[(\partial_{\tau}n^a)^2+v_s^2(\partial_{r}n^a)^2].
\label{1.4} \end{eqnarray}
Three phenomenological parameters enter into Eq.(\ref{1.4}):
the parameter $\lambda$ represents a spin-fermion coupling, $g$ controls the strength of the antiferromagnetic (AF) fluctuations,
and $v_s$ stands for the spin velocity.

There are two basic distinctions between the conventional SF model \cite{chub} and that given by (\ref{1.4}).
i) The conventional SF model is controlled by a small parameter $\lambda/t\ll 1$. Only under this condition,
can the SF model be justified as a low-energy approximation to the Hubbard model. The physical meaning
of the coupling $\lambda$ in the conventional approach is that it represents a spin density wave (SDW) gap.
In contrast, in Eq.(\ref{1.4}), $\lambda$ should be considered as a coupling of order $t$  and it has nothing to do with the SDW gap.

ii) The spin sector in the conventional SF model is described by a soft spin mode $\vec\phi$ which appears as a Hubbard-Stratonovich auxiliary field to linearize the quartic fermion coupling. This mode is "made out" of the fermionic degrees of freedom which necessarily implies that the spin velocity $v_s$ is of order the relevant Fermi velocity, $v_s\sim v_F\sim ta$, with $a$ being
the lattice spacing.
In contrast, in the model (\ref{1.4}), the spin sector is described instead by the "hard-spin" field $n^a$ that obeys the local constraint
$\sum_a(n^a)^2=1.$
The spin-$1$ order parameter $\vec n$ can be split into two spin-$1/2$
charge neutral spinons $z_{\alpha}:$
$$ n^a=\bar z_{i\alpha}\sigma^a_{\alpha\beta}z_{i\beta}, \quad \bar z_{i\alpha}z_{i\alpha}=1.$$
The emergent gauge field,
$z_{i\alpha}\to z_{i\alpha}e^{i\theta_i},$ glues them together. 

The compact $U(1)$ gauge theory is at strong coupling (there is no Maxwell term in the action) and it is presumably  in a confining phase. 
If one assumes however that there exists an energy window in which the deconfinement does indeed take place, 
one can derive the small-pocket FS  
within an  effective FL$^*$ theory based on the spinon decomposition of the order parameter and on symmetry considerations.\cite{sachdev}  
It is not totally clear what controls such a theory, however.
For instance,  the key equation (3.1) in \cite{sachdev} rewrites the conventional electron on-site operator 
$c_{\alpha}$  in terms of a new (unitary equivalent ) $SU(2)$ spinor $\psi_{\pm}$:
\begin{equation}
c_{\alpha}=z_{\alpha}\psi_+ -\epsilon _{\alpha\beta}\bar z_{\beta}\psi_-
\approx Z(F_{\alpha}+G_{\alpha}),
\label{spinor}
\end{equation} 
where $Z$ is some constant factor, $\epsilon _{\alpha\beta}$ is a totally antisymmetric tensor, $\epsilon _{\uparrow\downarrow}=1$, and $z_{\alpha}$ a set of complex numbers subject to the condition $\sum_{\alpha}\bar z_{\alpha}z_{\alpha}=1$.  
Although the electron operator $c_{\alpha}$ by definition transforms itself as  an $SU(2)$ spinor, the $z_{\alpha}\psi_+$ and $\epsilon _{\alpha\beta}\bar z_{\beta}\psi_-$ operators taken apart from each other do not
transform in the same way. Nevertheless, a true single fermion operator splits into a sum of two {\it independent} fermion modes, $F_{\alpha}$ and $G_{\alpha}$. It is not obvious what controls the approximation made in Eq.(\ref{spinor}).

On the other hand, when an emergent gauge field is indeed in 
the deconfined phase (e.g., ${\cal Z}_2$ gauge field in $2+1$) the spinons can appear in the physical spectrum.
To get the gauge field
deconfined one needs first to  have it gapped. 
This is a necessary condition to arrive at the
FL* phase. One route to achieve this is to apply a MF theory that breaks $U(1)$ down to  
${\cal Z}_2$ \cite{sachdev1}. However, such a theory can be controlled only in the limit
where the spin exchange term is much larger than  the hopping term which is not the case for the cuprates, however.  

An alternative approach  was recently proposed in \cite{sachdev2}.
Roughly speaking it involves a direct coupling of a spinon field in the phenomenological SF model (\ref{1.4}) to a ${\cal Z}_2$  gauge field. A crucial requirement in this approach is that the coupling $\lambda$ is to be considered as the largest energy scale in the problem. The topological order and fractionalization appear in a regime inaccessible in a small $\lambda$ expansion. 

\section{III. Itinerant-localized model of strong correlations}

In the present paper we show that the starting point employed in \cite{sachdev} -- the phenomenological SF model (\ref{1.4}) -- can in fact be derived microscopically up to a certain $\lambda$-dependent extra term by employing
the earlier established mapping of the $t-J$ model of strongly correlated electrons onto the Kondo-Heisenberg model at
a dominantly large Kondo coupling \cite{pfk}. This enable us to clarify the physical meaning of the coupling $\lambda$  in (\ref{1.4}) and, in this way, to understand the possible physical limitations of the SF models in describing the physics of the  underdoped cuprates.

Let us start with the observation that,
in the underdoped cuprates, one striking feature is the simultaneous existence of both
localized and itinerant nature of the lattice electrons.
To include both aspects into consideration on equal footing,
Ribeiro and Wen proposed a slave-particle spin-dopon representation of the projected electron operators
in the enlarged Hilbert space \cite{wen},
\begin{eqnarray}
\tilde c_{i}^{\dagger}
=c_{i}^{\dagger}(1-n_{i-\sigma})=\frac{1}{\sqrt{2}}(\frac{1}{2}-2\vec S_i\cdot\vec\sigma)\tilde d_i.
\label{tildec}\end{eqnarray}
In this framework, the localized electron is represented by the lattice spin $\vec S\in su(2)$
whereas  the doped hole (dopon) is described by the projected hole operator,
$\tilde{d}_{i\alpha}=d_{i\sigma}(1-n^d_{i-\alpha})$.
Here $\tilde c^{\dagger}=(\tilde c^{\dagger}_{\uparrow},\tilde c^{\dagger}_{\downarrow})^{t}$ and
$\tilde d=(\tilde d_{\uparrow}, \tilde d_{\downarrow})^{t}$.

In terms of the projected electron operators, the constraint of no double occupancy (NDO) takes on the form
\begin{equation}
\sum_{\alpha}(\tilde c_{i\alpha}^{\dagger}\tilde c_{i\alpha})+\tilde c_{i\alpha}\tilde c_{i\alpha}^{\dagger}=1.
\label{a} \end{equation}
It singles out the physical $3d$ on-site Hilbert space. Only under  this condition are the projected electron operators
isomorphic to the original Hubbard operators.
Within the spin-dopon representation, the NDO constraint reduces to a Kondo-type interaction,\cite{pfk}
\begin{equation}
\vec{S_i} \cdot
\vec{s_i}+\frac{3}{4}(\tilde{d}_{i\uparrow}^{\dagger}\tilde{d}_{i\uparrow}+
\tilde{d}_{i\downarrow}^{\dagger}\tilde{d}_{i\downarrow})=0,
\label{cnstr}\end{equation}
with $\vec
s_i=\sum_{\alpha,\beta}\tilde{d}_{i\alpha}^{\dagger}\vec\sigma_{\alpha\beta}\tilde{d}_{i\beta}
$  being the dopon spin operator.

Proceeding with Eq.(\ref{cnstr}) we end up with the lattice Kondo-Heisenberg  model which,
at strong coupling $(\lambda\gg t)$, is  equivalent to
the original $t-J$ model:
\begin{eqnarray}
H_{t-J}&=& \sum_{ij\sigma} 2t_{ij}
{d}_{i\sigma}^{\dagger} {d}_{j\sigma}
+ J\sum_{ij} \vec S_i(1-n_i^d) \cdot \vec S_j(1-n_j^d)\nonumber\\
&+& \lambda
\sum_i(\vec{S_i} \cdot
\vec{s_i}+\frac{3}{4}{n}^d_i),\quad \lambda\to +\infty.
\label{2.7}
\end{eqnarray}
In spite of the global character of the parameter $\lambda$, it  enforces the NDO constraint locally due to the fact that
the on-site physical Hilbert subspace corresponds to zero eigenvalues of the constraint, whereas the nonphysical subspace
is spanned by the eigenvectors with strictly positive eigenvalues. As we showed elsewhere, in $1d$, Eq. (\ref{2.7}) reproduces
the well-known exact results for the $t-J$ model.\cite{fk}

The unphysical doubly occupied electron states are separated from the physical
sector by an energy gap $\sim\lambda$.
In the $\lambda\to +\infty$ limit, i.e. in the limit in which $\lambda$ is much larger than any other existing energy
scale in the problem, those states
are  automatically excluded from the Hilbert space.
On the other hand, in this limit, the high and low energy itinerant fermions cannot be separated out and this is another
manifestation of the local Mott physics.
In view of that it does not seem appropriate to integrate out the high-energy fermions. We assume however that this separation still holds
in the spin-dopon representation for well localized
particles, i.e., the localized spin degrees of freedom.
In this way we arrive \cite{fk1} at an
effective Lagrangian to describe strongly correlated electrons 
in the underdoped phase:
$$\tilde{\cal L}=\tilde{\cal L}_{d}+\tilde{\cal L}_{\lambda}+\tilde{\cal L}_{n},$$
\begin{eqnarray}
\tilde{\cal L}_{d}&=&\sum_i d^{\dagger}_{i\alpha}(\partial_{\tau}
+\frac{3\lambda}{4}-\mu)d_{i\alpha}\nonumber\\
&+&2\sum_{i<j}t_{ij}(d^{\dagger}_{i\alpha}d_{j\alpha}+
d^{\dagger}_{j\alpha}d_{i\alpha}),\nonumber\\
\tilde{\cal L}_{\lambda}&=&\lambda\sum_i(-1)^{x_i+y_i}n_i^{a} d^{\dagger}_{i\alpha}\sigma^a_{\alpha\beta}d_{i\beta},\nonumber\\
\tilde{\cal L}_{n}&=&\frac{1}{2g}\int d^2\vec r[(\partial_{\tau}n^a)^2+v_s^2(\partial_{\vec r}n^a)^2].
\label{1.22} \end{eqnarray}
where $v_s=Ja$ is the spin-wave velocity, and $g=  4Ja^2/s$.
In this model, the itinerant degrees of freedom are described by the dopons whereas the localized ones -- by the fluctuating AF order parameter, $n^a$.  In the limit $\lambda\to \infty$, 
the dopons and the lattice spins merge into the gauge neutral objects (\ref{tildec}) to represent the physical electrons.

There are a few distinctions between ${\cal L'}$ and $\tilde{\cal L}$.
To start with, the model $\tilde{\cal L}$ is formulated in terms of the hole (dopon) operators $d's$, rather than
the electron operators $c's$. It might seem, at first sight, that the electron-like pockets should emerge in place of the
hole-like pockets. However, this is not the case, since in our model the fermion dispersion changes sign, $t_{ij}
\to -t_{ij}$. Note also that
the renormalized hopping amplitude in $\tilde{\cal L}$ contains the spin-fermion coupling $\lambda$, as given by  Eq.(\ref{1.22}).
Because of this, in the physical limit, $\lambda\to +\infty$, the model remains well defined and finite.
Finally, the hopping amplitude in $\tilde{\cal L}$ contains an important factor of $2$ that reflects the fact that a vacancy is now represented as a spin-dopon singlet (see, \cite{pfk}).

The physical meaning of the Kondo coupling $\lambda$ is that it takes care of the NDO constraint. 
As the quantum Monte Carlo simulations of the model (\ref{2.7}) explicitly show \cite{ilya}, the NDO constraint
plays the dominating role in the destruction of the long-range AF state. The critical hole concentration is
around  $x_c=0.05$ at $J=0.2t.$ The spin-spin correlation functions at fixed doping become almost identical to each other
for $\lambda\ge 10t$. This indicates that, in principle,
finite but large enough values of $\lambda$ already provide a reliable description of the existing strong correlations in the underdoped regime.

\section{IV. Further comparisons of the models}

With this understanding of the physical meaning of the Kondo coupling $\lambda$, 
let us compare the ${\cal L}$ and ${\cal L'}$ SF models against the itinerant-localized $\tilde{\cal L}$  model.  As the conventional SF model (\ref{1.1}) implies $\lambda/t\ll 1$, it does not capture strong correlations at all. Its phenomenological modification given by (\ref{1.4}) implies that $\lambda \sim t$. This might correspond to an optimally doped phase, provided the extra $\lambda$-term in the hopping term is also taken into account. In fact, its magnitude is of the same order as that of the $t$-term. 
In the ${\cal Z}_2$ gauge theory based on ${\cal L'}$ and described in detail in \citep{sachdev2}, the crucial statement is that the $\lambda$ is the largest energy scale.
This, by definition, implies that the theory must remain finite and well defined at $\lambda =+\infty,$ which physically corresponds to a regime of strong electron correlations.   

There is an analogy with  that and the conventional large $U$ approximation to the Hubbard model referred to as the $t-J$ model. Namely, the Hubbard model reduces at $U=\infty$ to a well defined model that  describes the so-called Nagaoka phase of strongly correlated electrons.  The ground-state energy is finite and independent of $U$ in the large-$U$ limit.
In contrast, the states obtained in \cite{sachdev2} in the leading in 
$\lambda$ approximation do not exhibit such a behaviour. For example,
the ground state of the Mott insulator $(x=0)$ reads
$E_G=-\lambda +\Delta -\mu$. The energy of the spinon excitation above the Mott insulator state takes the form 
$E_s=-\mu +2\Delta -\sqrt{\lambda^2+4\Delta^2}$. Here $\mu$ and $\Delta$ are certain asymptotically $\lambda$-independent parameters. In particular, the FL$^*$ state is favoured in the limit $\lambda\gg \Delta\gg |t|\gg |J|$. \cite{sachdev2}
This implies that the Mott insulator is not a stable phase in the physically well defined regime of 
strong electron correlations in which $\lambda$ becomes boundless. This is hardly appropriate since  those correlations are  
in fact a key ingredient behind the Mott physics. 

On the other hand, if we follow all the steps exposed in \cite{sachdev2}
to compute those quantities starting right from the itinerant-localized model (\ref{1.22}) 
we will obtain instead
$E_G=\Delta -\mu$ and 
$E_s=-\mu +\lambda/2 +2\Delta -\sqrt{\lambda^2/4+4\Delta^2}.$
Both expressions remain finite, in the large $\lambda$ limit.  
This occurs due to the fact that model (\ref{1.22}) involves the extra $\lambda$-dependent renormalization of the hopping term which is missed in both ${\cal L}$ and ${\cal L'}$ 
SF models.

\section{V. Conclusion}

To conclude, we show that a number of unsatisfactory points that arise in an attempt to apply the  
SF-like  models to treat the underdoped cuprates can be safely removed provided the itinerant-localized model of strongly correlated electrons is employed in their place. This indicates that this model may be the correct starting point to deal with the strong correlation physics displayed by the underdoped cuprates. Within a framework of the quantum Monte Carlo simulations, this model has recently been shown to provide a few interesting new results. In particular, the strong electron correlations are shown to play a key role in the abrupt destruction of the quasi-long-range AF order in the lightly doped regime.\cite{ilya} This model was also employed to study the Nagaoka ($U=\infty$) limit of the Hubbard model to 
demonstrate the absence of the long-range ferromagnetic ordering  at finite doping.  
\cite{ilya2} 

A further possible application of the itinerant-localized model might be to theoretically explore an experimentally observed instability towards a d-wave formation of a charge order in the PG phase. This charge order seems to compete directly with the d-wave superconducting regime. 
There is a strong evidence that the observed charge order is due to strong electron correlations. 
This work is already in progress and results will be presented elsewhere.

\end{document}